\documentclass[screen=true, sigplan]{acmart}
\usepackage[utf8]{inputenc}
\usepackage[T1]{fontenc}

\newcommand{\packageGraphicx}{\usepackage{graphicx}}
\newcommand{\packageHyperref}{\usepackage{hyperref}}
\newcommand{\renewrmdefault}{\renewcommand{\rmdefault}{ptm}}
\newcommand{\packageRelsize}{\usepackage{relsize}}
\newcommand{\packageAmsmath}{\usepackage{amsmath}}
\newcommand{\packageMathabx}{\usepackage{mathabx}}
\newcommand{\packageWasysym}{
  \let\leftmoon\relax \let\rightmoon\relax \let\fullmoon\relax \let\newmoon\relax \let\diameter\relax
  \usepackage[nointegrals]{wasysym}}
\newcommand{\packageTxfonts}{
  \let\widering\relax
  \let\oldwidebar\widebar
  \let\widebar\relax
  \usepackage{newtxmath}
  \ifx\widebar\relax
    \let\widebar\oldwidebar
  \fi
}
\newcommand{\packageTextcomp}{\usepackage{textcomp}}
\newcommand{\packageFramed}{\usepackage{framed}}
\newcommand{\packageHyphenat}{\usepackage[htt]{hyphenat}}
\newcommand{\packageColor}{\usepackage[usenames,dvipsnames]{color}}
\newcommand{\doHypersetup}{\hypersetup{bookmarks=true,bookmarksopen=true,bookmarksnumbered=true}}
\newcommand{\packageTocstyle}{}
\newcommand{\packageCJK}{\IfFileExists{CJK.sty}{\usepackage{CJK}}{}}
\renewcommand\packageColor\relax
\renewcommand\packageTocstyle\relax
\renewcommand\packageMathabx{\ifx\bigtimes\undefined \usepackage{mathabx} \else \relax \fi}
\renewcommand\packageTxfonts\relax

\renewcommand{\renewrmdefault}{}

\packageGraphicx
\packageHyperref
\renewrmdefault
\packageRelsize
\packageAmsmath
\packageMathabx
\packageWasysym
\packageTxfonts
\packageTextcomp
\packageFramed
\packageHyphenat
\packageColor
\doHypersetup
\packageTocstyle
\packageCJK

\newcommand{\sectionNewpage}{}

\newcommand{\preDoc}{}
\newcommand{\postDoc}{}

\newcommand{\BookRef}[2]{\emph{#2}}
\newcommand{\ChapRef}[2]{\SecRef{#1}{#2}}
\newcommand{\SecRef}[2]{section~#1}
\newcommand{\PartRef}[2]{part~#1}
\newcommand{\BookRefUC}[2]{\BookRef{#1}{#2}}
\newcommand{\ChapRefUC}[2]{\SecRefUC{#1}{#2}}
\newcommand{\SecRefUC}[2]{Section~#1}
\newcommand{\PartRefUC}[2]{Part~#1}

\newcommand{\BookRefLocal}[3]{\hyperref[#1]{\BookRef{#2}{#3}}}
\newcommand{\ChapRefLocal}[3]{\hyperref[#1]{\ChapRef{#2}{#3}}}
\newcommand{\SecRefLocal}[3]{\hyperref[#1]{\SecRef{#2}{#3}}}
\newcommand{\PartRefLocal}[3]{\hyperref[#1]{\PartRef{#2}{#3}}}
\newcommand{\BookRefLocalUC}[3]{\hyperref[#1]{\BookRefUC{#2}{#3}}}
\newcommand{\ChapRefLocalUC}[3]{\hyperref[#1]{\ChapRefUC{#2}{#3}}}
\newcommand{\SecRefLocalUC}[3]{\hyperref[#1]{\SecRefUC{#2}{#3}}}
\newcommand{\PartRefLocalUC}[3]{\hyperref[#1]{\PartRefUC{#2}{#3}}}

\newcommand{\BookRefUN}[1]{\BookRef{}{#1}}

\newcommand{\SecRefUN}[1]{``#1''}

\newcommand{\BookRefLocalUN}[2]{\hyperref[#1]{\BookRefUN{#2}}}

\newcommand{\SecRefLocalUN}[2]{\hyperref[#1]{\SecRefUN{#2}}}

\newcommand{\SectionNumberLink}[2]{\hyperref[#1]{#2}}

\newcommand{\Scribtexttt}[1]{{\texttt{#1}}}

\newcommand{\planetName}[1]{PLane\hspace{-0.1ex}T}

\newcommand{\Stttextmore}{{\fontencoding{T1}\selectfont>}}
\newcommand{\Stttextless}{{\fontencoding{T1}\selectfont<}}

\makeatletter
\def\empty@finalstrut#1{%
  \unskip\ifhmode\nobreak\fi\vrule\@width\z@\@height\z@\@depth\z@}
\def\no@strut{\global\setbox\@arstrutbox\hbox{%
    \vrule \@height\z@
           \@depth\z@
           \@width\z@}%
    \gdef\@endpbox{\empty@finalstrut\@arstrutbox\par\egroup\hfil}%
}%
\def\yes@strut{\global\setbox\@arstrutbox\hbox{%
    \vrule \@height\arraystretch \ht\strutbox
           \@depth\arraystretch \dp\strutbox
           \@width\z@}%
    \gdef\@endpbox{\@finalstrut\@arstrutbox\par\egroup\hfil}%
}%
\def\@mkpream#1{\@firstamptrue\@lastchclass6
  \let\@preamble\@empty\def\empty@preamble{\add@ins}%
  \let\protect\@unexpandable@protect
  \let\@sharp\relax\let\add@ins\relax
  \let\@startpbox\relax\let\@endpbox\relax
  \@expast{#1}%
  \expandafter\@tfor \expandafter
    \@nextchar \expandafter:\expandafter=\reserved@a\do
       {\@testpach\@nextchar
    \ifcase \@chclass \@classz \or \@classi \or \@classii \or \@classiii
      \or \@classiv \or\@classv \fi\@lastchclass\@chclass}%
  \ifcase \@lastchclass \@acol
      \or \or \@preamerr \@ne\or \@preamerr \tw@\or \or \@acol \fi}
\def\@addamp{%
  \if@firstamp
    \@firstampfalse
    \edef\empty@preamble{\add@ins}%
  \else
    \edef\@preamble{\@preamble &}%
    \edef\empty@preamble{\expandafter\noexpand\empty@preamble &\add@ins}%
  \fi}
\newif\iftw@hlines \tw@hlinesfalse
\def\@xhline{\ifx\reserved@a\hline
               \tw@hlinestrue
             \else\ifx\reserved@a\Hline
               \tw@hlinestrue
             \else
               \tw@hlinesfalse
             \fi\fi
      \iftw@hlines
        \aftergroup\do@after
      \fi
      \ifnum0=`{\fi}%
}
\def\do@after{\emptyrow[\the\doublerulesep]}
\def\emptyrow{\noalign\bgroup\@ifnextchar[\@emptyrow{\@emptyrow[\z@]}}
\def\@emptyrow[#1]{\no@strut\gdef\add@ins{\vrule \@height\z@ \@depth#1 \@width\z@}\egroup%
\empty@preamble\\
\noalign{\yes@strut\gdef\add@ins{\vrule \@height\z@ \@depth\z@ \@width\z@}}%
}
\def\tabrow#1{\noalign\bgroup\@ifnextchar[{\@tabrow{#1}}{\@tabrow{#1}[]}}
\def\@tabrow#1[#2]{\no@strut\egroup#1\ifx.#2.\\\else\\[#2]\fi\noalign{\yes@strut}}
\def\endpltstabular{\crcr\egroup\egroup \egroup}
\expandafter \let \csname endpltstabular*\endcsname = \endpltstabular
\def\pltstabular{\let\@halignto\@empty\@pltstabular}
\@namedef{pltstabular*}#1{\def\@halignto{to#1}\@pltstabular}
\def\@pltstabular{\leavevmode \bgroup \let\@acol\@tabacol
   \let\@classz\@tabclassz
   \let\@classiv\@tabclassiv \let\\\@tabularcr\@stabarray}
\def\@stabarray{\m@th\@ifnextchar[\@sarray{\@sarray[c]}}
\def\@sarray[#1]#2{%
  \bgroup
  \setbox\@arstrutbox\hbox{%
    \vrule \@height\arraystretch\ht\strutbox
           \@depth\arraystretch \dp\strutbox
           \@width\z@}%
  \@mkpream{#2}%
  \edef\@preamble{%
    \ialign \noexpand\@halignto
      \bgroup \@arstrut \@preamble \tabskip\z@skip \cr}%
  \let\@startpbox\@@startpbox \let\@endpbox\@@endpbox
  \let\tabularnewline\\%
    \let\@sharp##%
    \set@typeset@protect
    \lineskip\z@skip\baselineskip\z@skip
    \@preamble}

\makeatother

\newenvironment{bigtabular}{\begin{pltstabular}}{\end{pltstabular}}

\newlength{\stabLeft}
\newcommand{\bigtableleftpad}{\hspace{\stabLeft}}

\newenvironment{SingleColumn}{\begin{list}{}{\topsep=0pt\partopsep=0pt%
\listparindent=0pt\itemindent=0pt\labelwidth=0pt\leftmargin=0pt\rightmargin=0pt%
\itemsep=0pt\parsep=0pt}\item}{\end{list}}

\newcommand{\SCodePreSkip}{\vskip\abovedisplayskip}
\newcommand{\SCodePostSkip}{\vskip\belowdisplayskip}

\newcommand{\SVInsetPreSkip}{\vskip\abovedisplayskip}
\newcommand{\SVInsetPostSkip}{\vskip\belowdisplayskip}

\newcommand{\titleAndVersionAndAuthors}[3]{\title{#1\\{\normalsize \SVersionBefore{}#2}}\author{#3}\maketitle}

\newcommand{\titleAndEmptyVersionAndAuthors}[3]{\title{#1}\author{#3}\maketitle}

\newcommand{\SAuthor}[1]{#1}
\newcommand{\SAuthorSep}[1]{\qquad}
\newcommand{\SVersionBefore}[1]{Version }

\newcommand{\SNumberOfAuthors}[1]{}

\let\SOriginalthesubsection\thesubsection
\let\SOriginalthesubsubsection\thesubsubsection

\newcommand{\Ssection}[2]{\section[#1]{#2}\let\thesubsection\SOriginalthesubsection}
\newcommand{\Ssubsection}[2]{\subsection[#1]{#2}\let\thesubsubsection\SOriginalthesubsubsection}

\newcommand{\Ssectionstar}[1]{\section*{#1}\renewcommand*\thesubsection{\arabic{subsection}}\setcounter{subsection}{0}}

\newcommand{\Ssectionstarx}[2]{\Ssectionstar{#2}\phantomsection\addcontentsline{toc}{section}{#1}}

\newcounter{GrouperTemp}

\newenvironment{SVerbatim}{}{}

\newcommand{\Snolinkurl}[1]{\nolinkurl{#1}}

\newcommand{\SAuthorinfo}[4]{#1}
\newcommand{\SAuthorPlace}[1]{#1}
\newcommand{\SAuthorEmail}[1]{#1}
\newcommand{\SAuthorOrcid}[1]{#1}

\newcommand{\SConferenceInfo}[2]{}
\newcommand{\SCopyrightYear}[1]{}
\newcommand{\SCopyrightData}[1]{}
\newcommand{\Sdoi}[1]{}

\newcommand{\SCategory}[3]{}
\newcommand{\SCategoryPlus}[4]{}
\newcommand{\STerms}[1]{}
\newcommand{\SKeywords}[1]{}

\newcommand{\AutobibLink}[1]{\color{ACMPurple}{#1}}
\setcopyright{acmlicensed}
\acmDOI{10.1145/3677998.3678223}
\acmYear{2024}
\copyrightyear{2024}
\acmISBN{979-8-4007-1101-5/24/09}
\acmConference[FUNARCH '24]{Proceedings of the 2nd ACM SIGPLAN International Workshop on Functional Software Architecture}{September 6, 2024}{Milan, Italy}
\acmBooktitle{Proceedings of the 2nd ACM SIGPLAN International Workshop on Functional Software Architecture (FUNARCH '24), September 6, 2024, Milan, Italy}
\acmSubmissionID{icfpws24funarchmain-id1-p}
\received{2024-06-03}
\received[accepted]{2024-06-30}

\newcommand{\Autobibtarget}[1]{\phantomsection#1}

\usepackage{calc}
\newlength{\ABcollength}
\newcommand{\Autocolbibnumber}[1]{\parbox[t]{5ex}{\hfill#1~~\vspace{1.0ex}}}
\newcommand{\Autocolbibentry}[1]{\setlength{\ABcollength}{\linewidth-5ex}\parbox[t]{\ABcollength}{#1\vspace{1.0ex}}}

\newcommand{\Autobibref}[1]{#1}

\providecommand{\AutobibLink}[1]{#1}

\newcommand{\pseudodoi}[1]{#1}

\newcommand{\NoteBox}[1]{\footnote{#1}}
\newcommand{\NoteContent}[1]{#1}

\newcommand{\FootnoteRef}[1]{}
\newcommand{\FootnoteTarget}[1]{}

\newcommand{\FootnoteBlockContent}[1]{}
\usepackage{ccaption}

\makeatletter
\@ifundefined{belowcaptionskip}{}{}
\makeatother

\newcommand{\Legend}[1]{~

                        \hrule width \hsize height .33pt
                        \vspace{4pt}
                        \legend{#1}}

\newcommand{\FigureTarget}[2]{#1}
\newcommand{\FigureRef}[2]{#1}

\newlength{\FigOrigskip}
\FigOrigskip=\parskip

\newcommand{\FigureSetRef}{\refstepcounter{figure}}

\newenvironment{FigureMulti}{\begin{figure*}[t!p]\FigureSetRef}{\end{figure*}}

\newenvironment{Herefigure}{\begin{figure}[ht!]\FigureSetRef\centering}{\end{figure}}

\newenvironment{Centerfigure}{\begin{Xfigure}\centering\item}{\end{Xfigure}}
\newenvironment{Leftfigure}{\begin{Xfigure}\item}{\end{Xfigure}}

\newenvironment{Xfigure}{\begin{list}{}{\leftmargin=0pt\topsep=0pt\parsep=\FigOrigskip\partopsep=0pt}}{\end{list}}

\newenvironment{FigureInside}{}{}

\newcommand{\Centertext}[1]{\begin{center}#1\end{center}}

\newcommand{\SColorize}[2]{\color{#1}{#2}}

\newcommand{\SHyphen}[1]{#1}

\newcommand{\inColor}[2]{{\SHyphen{\Scribtexttt{\SColorize{#1}{#2}}}}}
\definecolor{PaleBlue}{rgb}{0.90,0.90,1.0}
\definecolor{LightGray}{rgb}{0.90,0.90,0.90}
\definecolor{CommentColor}{rgb}{0.76,0.45,0.12}
\definecolor{ParenColor}{rgb}{0.52,0.24,0.14}
\definecolor{IdentifierColor}{rgb}{0.15,0.15,0.50}
\definecolor{ResultColor}{rgb}{0.0,0.0,0.69}
\definecolor{ValueColor}{rgb}{0.13,0.55,0.13}
\definecolor{OutputColor}{rgb}{0.59,0.00,0.59}

\newcommand{\RktPn}[1]{\inColor{ParenColor}{#1}}

\newcommand{\RktSym}[1]{\inColor{IdentifierColor}{#1}}

\newcommand{\RktVal}[1]{\inColor{ValueColor}{#1}}

\newcommand{\RktRdr}[1]{\inColor{black}{#1}}
\newcommand{\RktVarCol}[1]{\inColor{IdentifierColor}{#1}}
\newcommand{\RktVar}[1]{{\RktVarCol{\textsl{#1}}}}

\newenvironment{RktBlk}{}{}

\newcommand{\RBackgroundLabel}[1]{}

\renewcommand{\titleAndVersionAndAuthors}[3]{\title{#1}#3\maketitle}
\renewcommand{\titleAndEmptyVersionAndAuthors}[3]{\titleAndVersionAndAuthors{#1}{#2}{#3}}

\def\SAuthor#1{\SAutoAuthor#1\SAutoAuthorDone{#1}}
\def\SAutoAuthorDone#1{}
\def\SAutoAuthor{\futurelet\next\SAutoAuthorX}
\def\SAutoAuthorX{\ifx\next\SAuthorinfo \let\Snext\relax \else \let\Snext\SToAuthorDone \fi \Snext}
\def\SToAuthorDone{\futurelet\next\SToAuthorDoneX}
\def\SToAuthorDoneX#1{\ifx\next\SAutoAuthorDone \let\Snext\SAddAuthorInfo \else \let\Snext\SToAuthorDone \fi \Snext}
\newcommand{\SAddAuthorInfo}[1]{\SAuthorinfo{#1}{}{}}

\renewcommand{\SAuthorinfo}[4]{\author{#1}{#2}{#3}{#4}}
\renewcommand{\SAuthorSep}[1]{}
\renewcommand{\SAuthorOrcid}[1]{\orcid{#1}}
\renewcommand{\SAuthorPlace}[1]{\affiliation{#1}}
\renewcommand{\SAuthorEmail}[1]{\email{#1}}

\renewcommand{\SConferenceInfo}[2]{\conferenceinfo{#1}{#2}}
\renewcommand{\SCopyrightData}[1]{\copyrightdata{#1}}

\renewcommand{\SCategory}[3]{\category{#1}{#2}{#3}}
\renewcommand{\SCategoryPlus}[4]{\category{#1}{#2}{#3}[#4]}
\renewcommand{\STerms}[1]{\terms{#1}}
\renewcommand{\SKeywords}[1]{\keywords{#1}}

\newcommand{\SccsdescNumber}[2]{\ccsdesc[#1]{#2}}
\renewcommand{\Legend}[1]{\legend{#1}}
\begin{document}
\preDoc

\begin{abstract}Surveys and experiments in economics involve stateful interactions:
participants receive different messages based on earlier answers,
choices, and performance, or trade across many rounds with other
participants. In the design of Congame, a platform for running such
economic studies, we decided to use delimited continuations to manage
the common flow of participants through a study. Here we report on
the positives of this approach, as well as some challenges of using
continuations, such as persisting data across requests, working with
dynamic variables, avoiding memory leaks, and the difficulty of
debugging continuations.\end{abstract}

\SccsdescNumber{500}{Software and its engineering~Reusability}

\SccsdescNumber{500}{Software and its engineering~Control structures}

\SccsdescNumber{300}{Software and its engineering~Domain specific languages}

\SccsdescNumber{300}{Software and its engineering~Data flow architectures}

\SccsdescNumber{300}{Software and its engineering~Functional languages}

\titleAndEmptyVersionAndAuthors{Continuations: What Have They Ever Done for Us? (Experience Report)}{}{\SNumberOfAuthors{2}\SAuthor{\SAuthorinfo{Marc Kaufmann}{\SAuthorOrcid{0000-0003-0366-6329}}{\SAuthorPlace{\institution{Central European University}\city{Vienna}\country{Austria}}}{\SAuthorEmail{kaufmannm@ceu.edu}}}\SAuthorSep{}\SAuthor{\SAuthorinfo{Bogdan Popa}{\SAuthorOrcid{0009-0001-7815-4742}}{\SAuthorPlace{\institution{Independent}\city{Cluj{-}Napoca}\country{Romania}}}{\SAuthorEmail{bogdan@defn.io}}}}
\label{t:x28part_x22Continuationsx5fx5fWhatx5fHavex5fTheyx5fEverx5fDonex5fforx5fUsx5fx5fx5fExperiencex5fReportx5fx22x29}

\noindent 

\noindent 

\noindent 

\noindent

\sectionNewpage

\Ssection{Introduction}{Introduction}\label{t:x28part_x22introx22x29}

Continuations in a web context allow applications to be programmed
in a direct style \Autobibref{~[\hyperref[t:x28autobib_x22Shriram_Krishnamurthix2c_Peter_Walton_Hopkinsx2c_Jay_McCarthyx2c_Paul_Tx2e_Graunkex2c_Greg_Pettyjohnx2c_and_Matthias_FelleisenImplementation_and_use_of_the_PLT_scheme_Web_serverHigherx2dOrder_and_Symbolic_Computation_20x2c_ppx2e_431x2dx2d4602007doix3a10x2e1007x2fs10990x2d007x2d9008x2dyx22x29]{\AutobibLink{5}}, \hyperref[t:x28autobib_x22Christian_QueinnecInverting_back_the_inversion_of_control_orx2c_continuations_versus_pagex2dcentric_programmingACM_SIGPLAN_Notices_38x2c_ppx2e_57x2dx2d642003doix3a10x2e1145x2f772970x2e772977x22x29]{\AutobibLink{7}}]}. In the design of
Congame, a platform for running economics studies, we took advantage
of this style of programming to implement a framework for specifying
composable surveys in a declarative way that elides most of the details
of day{-}to{-}day web programming from the study creator.

In particular, Congame automatically tracks and manages much of the
state of study participants, which is a big boon, since Congame studies
are inherently stateful applications. A participant{'}s next step may
depend on random treatments {---} as in A/B tests {---} or their own or
other participants{'} actions: they may only be allowed to move on if
they pass a comprehension test, and their payoff may be co{-}determined
by other participants with whom they interact in market games. Congame
thus frees the study creator from having to implement their own ad hoc
bug{-}ridden state management system.\NoteBox{\NoteContent{To riff on Greenspun{'}s Tenth Rule.}}

We report on our experience using delimited continuations to implement
Congame. In \ChapRef{\SectionNumberLink{t:x28part_x22minimalx22x29}{2}}{Mini Congame} we show a minimal implementation of a
system similar to Congame and demonstrate how natural it is to program
web applications in this style. Then, in \ChapRef{\SectionNumberLink{t:x28part_x22benefitsx22x29}{3}}{Benefits of Continuations} we note some
positive benefits that follow from our design and compare our system
to a popular platform for creating studies. In \ChapRef{\SectionNumberLink{t:x28part_x22challengesx22x29}{4}}{Challenges of Continuations} we
describe some challenges of managing the data flow and of debugging
in such a system. Finally, in \ChapRef{\SectionNumberLink{t:x28part_x22reflectionsx22x29}{5}}{Reflections \& Recommendations} we reflect on and
broadly recommend our approach, analyze its pros and cons, and make
suggestions targeted at other functional architects looking to implement
similar systems.

\sectionNewpage

\Ssection{Mini Congame}{Mini Congame}\label{t:x28part_x22minimalx22x29}

\begin{Herefigure}\begin{Centerfigure}\begin{FigureInside}\begin{RktBlk}\begin{SingleColumn}\RktPn{(}\RktSym{define}\mbox{\hphantom{\Scribtexttt{x}}}\RktSym{current{-}embed}\mbox{\hphantom{\Scribtexttt{x}}}\RktPn{(}\RktSym{make{-}parameter}\mbox{\hphantom{\Scribtexttt{x}}}\RktVal{\#f}\RktPn{)}\RktPn{)}

\RktPn{(}\RktSym{struct}\mbox{\hphantom{\Scribtexttt{x}}}\RktSym{step}\mbox{\hphantom{\Scribtexttt{x}}}\RktPn{(}\RktSym{handler}\RktPn{)}\RktPn{)}

\RktPn{(}\RktSym{struct}\mbox{\hphantom{\Scribtexttt{x}}}\RktSym{study}\mbox{\hphantom{\Scribtexttt{x}}}\RktPn{(}\RktSym{steps}\RktPn{)}\RktPn{)}

\RktPn{(}\RktSym{define}\mbox{\hphantom{\Scribtexttt{x}}}\RktPn{(}\RktSym{run{-}study}\mbox{\hphantom{\Scribtexttt{x}}}\RktSym{the{-}study}\RktPn{)}

\mbox{\hphantom{\Scribtexttt{xx}}}\RktPn{(}\RktSym{let}\mbox{\hphantom{\Scribtexttt{x}}}\RktSym{loop}\mbox{\hphantom{\Scribtexttt{x}}}\RktPn{(}\RktPn{[}\RktSym{steps}\mbox{\hphantom{\Scribtexttt{x}}}\RktPn{(}\RktSym{study{-}steps}\mbox{\hphantom{\Scribtexttt{x}}}\RktSym{the{-}study}\RktPn{)}\RktPn{]}\RktPn{)}

\mbox{\hphantom{\Scribtexttt{xxxx}}}\RktPn{(}\RktSym{if}\mbox{\hphantom{\Scribtexttt{x}}}\RktPn{(}\RktSym{null{\hbox{\texttt{?}}}}\mbox{\hphantom{\Scribtexttt{x}}}\RktSym{steps}\RktPn{)}

\mbox{\hphantom{\Scribtexttt{xxxxxxxx}}}\RktVal{{\textquotesingle}}\RktVal{(}\RktVal{continue}\RktVal{)}

\mbox{\hphantom{\Scribtexttt{xxxxxxxx}}}\RktPn{(}\RktSym{match}\mbox{\hphantom{\Scribtexttt{x}}}\RktPn{(}\RktSym{begin0}\mbox{\hphantom{\Scribtexttt{x}}}\RktPn{(}\RktSym{run{-}step}\mbox{\hphantom{\Scribtexttt{x}}}\RktPn{(}\RktSym{car}\mbox{\hphantom{\Scribtexttt{x}}}\RktSym{steps}\RktPn{)}\RktPn{)}

\mbox{\hphantom{\Scribtexttt{xxxxxxxxxxxxxxxxx}}}\RktPn{(}\RktSym{redirect/get/forget}\RktPn{)}\RktPn{)}

\mbox{\hphantom{\Scribtexttt{xxxxxxxxxx}}}\RktPn{[}\RktVal{{\textquotesingle}}\RktVal{(}\RktVal{retry}\RktVal{)}\mbox{\hphantom{\Scribtexttt{x}}}\RktPn{(}\RktSym{loop}\mbox{\hphantom{\Scribtexttt{x}}}\RktSym{steps}\RktPn{)}\RktPn{]}

\mbox{\hphantom{\Scribtexttt{xxxxxxxxxx}}}\RktPn{[}\RktVal{{\textquotesingle}}\RktVal{(}\RktVal{continue}\RktVal{)}\mbox{\hphantom{\Scribtexttt{x}}}\RktPn{(}\RktSym{loop}\mbox{\hphantom{\Scribtexttt{x}}}\RktPn{(}\RktSym{cdr}\mbox{\hphantom{\Scribtexttt{x}}}\RktSym{steps}\RktPn{)}\RktPn{)}\RktPn{]}\RktPn{)}\RktPn{)}\RktPn{)}\RktPn{)}

\RktPn{(}\RktSym{define}\mbox{\hphantom{\Scribtexttt{x}}}\RktPn{(}\RktSym{run{-}step}\mbox{\hphantom{\Scribtexttt{x}}}\RktSym{the{-}step}\RktPn{)}

\mbox{\hphantom{\Scribtexttt{xx}}}\RktPn{(}\RktSym{match}\mbox{\hphantom{\Scribtexttt{x}}}\RktSym{the{-}step}

\mbox{\hphantom{\Scribtexttt{xxxx}}}\RktPn{[}\RktPn{(}\RktSym{step}\mbox{\hphantom{\Scribtexttt{x}}}\RktPn{(}\RktSym{{\hbox{\texttt{?}}}}\mbox{\hphantom{\Scribtexttt{x}}}\RktSym{study{\hbox{\texttt{?}}}}\mbox{\hphantom{\Scribtexttt{x}}}\RktSym{substudy}\RktPn{)}\RktPn{)}

\mbox{\hphantom{\Scribtexttt{xxxxx}}}\RktPn{(}\RktSym{run{-}study}\mbox{\hphantom{\Scribtexttt{x}}}\RktSym{substudy}\RktPn{)}\RktPn{]}

\mbox{\hphantom{\Scribtexttt{xxxx}}}\RktPn{[}\RktPn{(}\RktSym{step}\mbox{\hphantom{\Scribtexttt{x}}}\RktSym{handler}\RktPn{)}

\mbox{\hphantom{\Scribtexttt{xxxxx}}}\RktPn{(}\RktSym{send/suspend/dispatch}

\mbox{\hphantom{\Scribtexttt{xxxxxx}}}\RktPn{(}\RktSym{lambda}\mbox{\hphantom{\Scribtexttt{x}}}\RktPn{(}\RktSym{embed}\RktPn{)}

\mbox{\hphantom{\Scribtexttt{xxxxxxxx}}}\RktPn{(}\RktSym{parameterize}\mbox{\hphantom{\Scribtexttt{x}}}\RktPn{(}\RktPn{[}\RktSym{current{-}embed}\mbox{\hphantom{\Scribtexttt{x}}}\RktSym{embed}\RktPn{]}\RktPn{)}

\mbox{\hphantom{\Scribtexttt{xxxxxxxxxx}}}\RktPn{(}\RktSym{response/xexpr}\mbox{\hphantom{\Scribtexttt{x}}}\RktPn{(}\RktSym{handler}\RktPn{)}\RktPn{)}\RktPn{)}\RktPn{)}\RktPn{)}\RktPn{]}\RktPn{)}\RktPn{)}\end{SingleColumn}\end{RktBlk}\end{FigureInside}\end{Centerfigure}

\Centertext{\Legend{\FigureTarget{\label{t:x28counter_x28x22figurex22_x22minimalx2d1x22x29x29}\textbf{Figure}~\textbf{1}. }{t:x28counter_x28x22figurex22_x22minimalx2d1x22x29x29}A mini Congame implementation.}}\end{Herefigure}

The core of Congame is a \emph{study}, represented as a tree of
\emph{steps} and sub{-}studies. Each \emph{step} in a study is a
procedure that generates a web page used to display and possibly
retrieve information to and from the participant being surveyed.
Figure~\hyperref[t:x28counter_x28x22figurex22_x22minimalx2d1x22x29x29]{\FigureRef{1}{t:x28counter_x28x22figurex22_x22minimalx2d1x22x29x29}} implements a minimal harness for constructing
and running these types of studies. A study creator uses the structures
defined in figure~\hyperref[t:x28counter_x28x22figurex22_x22minimalx2d1x22x29x29]{\FigureRef{1}{t:x28counter_x28x22figurex22_x22minimalx2d1x22x29x29}} alongside \emph{widgets} such as the
one defined in figure~\hyperref[t:x28counter_x28x22figurex22_x22minimalx2d2x22x29x29]{\FigureRef{2}{t:x28counter_x28x22figurex22_x22minimalx2d2x22x29x29}} to put together a study. The study
can then be run from within a Racket \Autobibref{~[\hyperref[t:x28autobib_x22Matthew_Flatt_and_PLTReferencex3a_RacketPLT_Design_Incx2ex2c_PLTx2dTRx2d2010x2d12010httpsx3ax2fx2fracketx2dlangx2eorgx2ftr1x2fx22x29]{\AutobibLink{3}}]} web server servlet
with \RktSym{run{-}study}.

\begin{Herefigure}\begin{Centerfigure}\begin{FigureInside}\begin{RktBlk}\begin{SingleColumn}\RktPn{(}\RktSym{define}\mbox{\hphantom{\Scribtexttt{x}}}\RktPn{(}\RktSym{button}\mbox{\hphantom{\Scribtexttt{x}}}\RktSym{label}\mbox{\hphantom{\Scribtexttt{x}}}\RktPn{[}\RktSym{action}\mbox{\hphantom{\Scribtexttt{x}}}\RktSym{void}\RktPn{]}\RktPn{)}

\mbox{\hphantom{\Scribtexttt{xx}}}\RktVal{{\textasciigrave}}\RktVal{(}\RktVal{a}

\mbox{\hphantom{\Scribtexttt{xxxx}}}\RktVal{(}\RktVal{[}\RktVal{href}\mbox{\hphantom{\Scribtexttt{x}}}\RktRdr{,}\RktPn{(}\RktPn{(}\RktSym{current{-}embed}\RktPn{)}

\mbox{\hphantom{\Scribtexttt{xxxxxxxxxxxxx}}}\RktPn{(}\RktSym{lambda}\mbox{\hphantom{\Scribtexttt{x}}}\RktPn{(}\RktVar{req}\RktPn{)}

\mbox{\hphantom{\Scribtexttt{xxxxxxxxxxxxxxx}}}\RktPn{(}\RktSym{action}\RktPn{)}

\mbox{\hphantom{\Scribtexttt{xxxxxxxxxxxxxxx}}}\RktVal{{\textquotesingle}}\RktVal{(}\RktVal{continue}\RktVal{)}\RktPn{)}\RktPn{)}\RktVal{]}\RktVal{)}

\mbox{\hphantom{\Scribtexttt{xxxx}}}\RktRdr{,}\RktSym{label}\RktVal{)}\RktPn{)}\end{SingleColumn}\end{RktBlk}\end{FigureInside}\end{Centerfigure}

\Centertext{\Legend{\FigureTarget{\label{t:x28counter_x28x22figurex22_x22minimalx2d2x22x29x29}\textbf{Figure}~\textbf{2}. }{t:x28counter_x28x22figurex22_x22minimalx2d2x22x29x29}A button {``}widget{''}.}}\end{Herefigure}

When a study is run, its steps are executed sequentially, and when a step uses a
widget, the widget reifies the current continuation of the step and stores it in
a hash table that maps URLs to continuations. The URL of that continuation is
then linked in the resulting HTML. Once a continuation URL is visited, the
continuation is restored so that \RktSym{run{-}step} returns and the study loop
can continue to the next step. Following this visit, the continuation is removed
from the hash table to prevent the participant from pressing the {``}Back{''} button
in their browser and redoing previous steps. The Racket Web Server
\Autobibref{~[\hyperref[t:x28autobib_x22Shriram_Krishnamurthix2c_Peter_Walton_Hopkinsx2c_Jay_McCarthyx2c_Paul_Tx2e_Graunkex2c_Greg_Pettyjohnx2c_and_Matthias_FelleisenImplementation_and_use_of_the_PLT_scheme_Web_serverHigherx2dOrder_and_Symbolic_Computation_20x2c_ppx2e_431x2dx2d4602007doix3a10x2e1007x2fs10990x2d007x2d9008x2dyx22x29]{\AutobibLink{5}}]} provides \RktSym{send/suspend/dispatch}, which takes care
of all the continuation management. Figure~\hyperref[t:x28counter_x28x22figurex22_x22examplex22x29x29]{\FigureRef{3}{t:x28counter_x28x22figurex22_x22examplex22x29x29}} shows a basic study
implemented using this framework.

\begin{Herefigure}\begin{Centerfigure}\begin{FigureInside}\begin{RktBlk}\begin{SingleColumn}\RktPn{(}\RktSym{define}\mbox{\hphantom{\Scribtexttt{x}}}\RktPn{(}\RktSym{hello}\RktPn{)}

\mbox{\hphantom{\Scribtexttt{xx}}}\RktVal{{\textasciigrave}}\RktVal{(}\RktVal{div}

\mbox{\hphantom{\Scribtexttt{xxxx}}}\RktVal{(}\RktVal{p}\mbox{\hphantom{\Scribtexttt{x}}}\RktVal{"Welcome to the study{\hbox{\texttt{.}}}"}\RktVal{)}

\mbox{\hphantom{\Scribtexttt{xxxx}}}\RktRdr{,}\RktPn{(}\RktSym{button}\mbox{\hphantom{\Scribtexttt{x}}}\RktVal{"Continue"}\RktPn{)}\RktVal{)}\RktPn{)}

\RktPn{(}\RktSym{define}\mbox{\hphantom{\Scribtexttt{x}}}\RktPn{(}\RktSym{done}\RktPn{)}

\mbox{\hphantom{\Scribtexttt{xx}}}\RktVal{{\textasciigrave}}\RktVal{(}\RktVal{p}\mbox{\hphantom{\Scribtexttt{x}}}\RktVal{"Thank you for participating{\hbox{\texttt{.}}}"}\RktVal{)}\RktPn{)}

\RktPn{(}\RktSym{define}\mbox{\hphantom{\Scribtexttt{x}}}\RktSym{example{-}study}

\mbox{\hphantom{\Scribtexttt{xx}}}\RktPn{(}\RktSym{study}

\mbox{\hphantom{\Scribtexttt{xxx}}}\RktPn{(}\RktSym{list}

\mbox{\hphantom{\Scribtexttt{xxxx}}}\RktPn{(}\RktSym{step}\mbox{\hphantom{\Scribtexttt{x}}}\RktSym{hello}\RktPn{)}

\mbox{\hphantom{\Scribtexttt{xxxx}}}\RktPn{(}\RktSym{step}\mbox{\hphantom{\Scribtexttt{x}}}\RktSym{done}\RktPn{)}\RktPn{)}\RktPn{)}\RktPn{)}\end{SingleColumn}\end{RktBlk}\end{FigureInside}\end{Centerfigure}

\Centertext{\Legend{\FigureTarget{\label{t:x28counter_x28x22figurex22_x22examplex22x29x29}\textbf{Figure}~\textbf{3}. }{t:x28counter_x28x22figurex22_x22examplex22x29x29}An example study.}}\end{Herefigure}

\sectionNewpage

\Ssection{Benefits of Continuations}{Benefits of Continuations}\label{t:x28part_x22benefitsx22x29}

Using continuations allows us to progress through the study by
traversing the study tree using regular techniques without having to
worry much about the fact that we are doing web programming. While
performing the traversal, we keep track of the participant{'}s position
in the tree which allows us to store participant data in a way that
imitates lexical scope, making it very natural for study writers to keep
track of local data.

Using continuations further allows us to use regular control flow
\Autobibref{~[\hyperref[t:x28autobib_x22Shriram_Krishnamurthix2c_Peter_Walton_Hopkinsx2c_Jay_McCarthyx2c_Paul_Tx2e_Graunkex2c_Greg_Pettyjohnx2c_and_Matthias_FelleisenImplementation_and_use_of_the_PLT_scheme_Web_serverHigherx2dOrder_and_Symbolic_Computation_20x2c_ppx2e_431x2dx2d4602007doix3a10x2e1007x2fs10990x2d007x2d9008x2dyx22x29]{\AutobibLink{5}}, \hyperref[t:x28autobib_x22Christian_QueinnecInverting_back_the_inversion_of_control_orx2c_continuations_versus_pagex2dcentric_programmingACM_SIGPLAN_Notices_38x2c_ppx2e_57x2dx2d642003doix3a10x2e1145x2f772970x2e772977x22x29]{\AutobibLink{7}}]}, meaning that every step of a study
can decide locally what the participant can do next. The actions in a
step can close over the step{'}s environment and use regular functional
programming techniques. Figure~\hyperref[t:x28counter_x28x22figurex22_x22varx2dexamplex22x29x29]{\FigureRef{4}{t:x28counter_x28x22figurex22_x22varx2dexamplex22x29x29}} illustrates that we
can write a step that tosses a coin and stores the outcome in the local
variable \RktSym{toss}, then offers a choice to the participant and,
after the participant makes their choice, checks the answer against
\RktSym{toss}.

\begin{Herefigure}\begin{Centerfigure}\begin{FigureInside}\begin{RktBlk}\begin{SingleColumn}\RktPn{(}\RktSym{defvar*}\mbox{\hphantom{\Scribtexttt{x}}}\RktSym{ok{\hbox{\texttt{?}}}}\RktPn{)}

\RktPn{(}\RktSym{defstep}\mbox{\hphantom{\Scribtexttt{x}}}\RktPn{(}\RktSym{intro}\RktPn{)}

\mbox{\hphantom{\Scribtexttt{xx}}}\RktPn{(}\RktSym{html}

\mbox{\hphantom{\Scribtexttt{xxx}}}\RktPn{(}\RktSym{h1}\mbox{\hphantom{\Scribtexttt{x}}}\RktVal{"Welcome to the study{\hbox{\texttt{!}}}"}\RktPn{)}

\mbox{\hphantom{\Scribtexttt{xxx}}}\RktPn{(}\RktSym{button}\mbox{\hphantom{\Scribtexttt{x}}}\RktVal{"Start"}\RktPn{)}\RktPn{)}\RktPn{)}

\RktPn{(}\RktSym{defstep}\mbox{\hphantom{\Scribtexttt{x}}}\RktPn{(}\RktSym{heads{-}or{-}tails}\RktPn{)}

\mbox{\hphantom{\Scribtexttt{xx}}}\RktPn{(}\RktSym{define}\mbox{\hphantom{\Scribtexttt{x}}}\RktSym{toss}\mbox{\hphantom{\Scribtexttt{x}}}\RktPn{(}\RktSym{random{-}ref}\mbox{\hphantom{\Scribtexttt{x}}}\RktVal{{\textquotesingle}}\RktVal{(}\RktVal{h}\mbox{\hphantom{\Scribtexttt{x}}}\RktVal{t}\RktVal{)}\RktPn{)}\RktPn{)}

\mbox{\hphantom{\Scribtexttt{xx}}}\RktPn{(}\RktSym{html}

\mbox{\hphantom{\Scribtexttt{xxx}}}\RktPn{(}\RktSym{button}

\mbox{\hphantom{\Scribtexttt{xxxx}}}\RktVal{"Heads"}

\mbox{\hphantom{\Scribtexttt{xxxx}}}\RktPn{(}\RktSym{$\lambda$}\mbox{\hphantom{\Scribtexttt{x}}}\RktPn{(}\RktPn{)}\mbox{\hphantom{\Scribtexttt{x}}}\RktPn{(}\RktSym{set{\hbox{\texttt{!}}}}\mbox{\hphantom{\Scribtexttt{x}}}\RktSym{ok{\hbox{\texttt{?}}}}\mbox{\hphantom{\Scribtexttt{x}}}\RktPn{(}\RktSym{eq{\hbox{\texttt{?}}}}\mbox{\hphantom{\Scribtexttt{x}}}\RktSym{toss}\mbox{\hphantom{\Scribtexttt{x}}}\RktVal{{\textquotesingle}}\RktVal{h}\RktPn{)}\RktPn{)}\RktPn{)}\RktPn{)}

\mbox{\hphantom{\Scribtexttt{xxx}}}\RktVal{" or "}

\mbox{\hphantom{\Scribtexttt{xxx}}}\RktPn{(}\RktSym{button}

\mbox{\hphantom{\Scribtexttt{xxxx}}}\RktVal{"Tails"}

\mbox{\hphantom{\Scribtexttt{xxxx}}}\RktPn{(}\RktSym{$\lambda$}\mbox{\hphantom{\Scribtexttt{x}}}\RktPn{(}\RktPn{)}\mbox{\hphantom{\Scribtexttt{x}}}\RktPn{(}\RktSym{set{\hbox{\texttt{!}}}}\mbox{\hphantom{\Scribtexttt{x}}}\RktSym{ok{\hbox{\texttt{?}}}}\mbox{\hphantom{\Scribtexttt{x}}}\RktPn{(}\RktSym{eq{\hbox{\texttt{?}}}}\mbox{\hphantom{\Scribtexttt{x}}}\RktSym{toss}\mbox{\hphantom{\Scribtexttt{x}}}\RktVal{{\textquotesingle}}\RktVal{t}\RktPn{)}\RktPn{)}\RktPn{)}\RktPn{)}\RktPn{)}\RktPn{)}

\RktPn{(}\RktSym{defstep}\mbox{\hphantom{\Scribtexttt{x}}}\RktPn{(}\RktSym{result}\RktPn{)}

\mbox{\hphantom{\Scribtexttt{xx}}}\RktPn{(}\RktSym{html}

\mbox{\hphantom{\Scribtexttt{xxx}}}\RktPn{(}\RktSym{if}\mbox{\hphantom{\Scribtexttt{x}}}\RktSym{ok{\hbox{\texttt{?}}}}

\mbox{\hphantom{\Scribtexttt{xxxxxxx}}}\RktPn{(}\RktSym{p}\mbox{\hphantom{\Scribtexttt{x}}}\RktVal{"You guessed right{\hbox{\texttt{.}}}"}\RktPn{)}

\mbox{\hphantom{\Scribtexttt{xxxxxxx}}}\RktPn{(}\RktSym{p}\mbox{\hphantom{\Scribtexttt{x}}}\RktVal{"You guessed wrong{\hbox{\texttt{.}}}"}\RktPn{)}\RktPn{)}\RktPn{)}\RktPn{)}

\RktPn{(}\RktSym{defstudy}\mbox{\hphantom{\Scribtexttt{x}}}\RktSym{choices}

\mbox{\hphantom{\Scribtexttt{xx}}}\RktPn{[}\RktSym{heads{-}or{-}tails}\mbox{\hphantom{\Scribtexttt{x}}}\RktSym{\mbox{{-}{-}{\Stttextmore}}}\mbox{\hphantom{\Scribtexttt{x}}}\RktRdr{,}\RktSym{end}\RktPn{]}\RktPn{)}

\RktPn{(}\RktSym{defstudy}\mbox{\hphantom{\Scribtexttt{x}}}\RktSym{example}

\mbox{\hphantom{\Scribtexttt{xx}}}\RktPn{[}\RktSym{intro}\mbox{\hphantom{\Scribtexttt{x}}}\RktSym{\mbox{{-}{-}{\Stttextmore}}}\mbox{\hphantom{\Scribtexttt{x}}}\RktSym{choices}\mbox{\hphantom{\Scribtexttt{x}}}\RktSym{\mbox{{-}{-}{\Stttextmore}}}\mbox{\hphantom{\Scribtexttt{x}}}\RktSym{result}\mbox{\hphantom{\Scribtexttt{x}}}\RktSym{\mbox{{-}{-}{\Stttextmore}}}\mbox{\hphantom{\Scribtexttt{x}}}\RktRdr{,}\RktSym{end}\RktPn{]}\RktPn{)}\end{SingleColumn}\end{RktBlk}\end{FigureInside}\end{Centerfigure}

\Centertext{\Legend{\FigureTarget{\label{t:x28counter_x28x22figurex22_x22varx2dexamplex22x29x29}\textbf{Figure}~\textbf{4}. }{t:x28counter_x28x22figurex22_x22varx2dexamplex22x29x29}A study showing lexical scope and data flow in Congame.}}\end{Herefigure}

Since our approach is data{-}driven, changing our data structures requires
only minor changes to our harness. For instance, adding support for
view handlers {---} study{-}specific static pages {---} meant extending the
\RktSym{step} struct with another field and adding one more request
handler to traverse the study tree and display those handlers as
necessary. If instead we had opted for a design where we store a
representation of steps in the database, then we would have had to
update the schema.

More generally, our design is flexible to changes. Extending studies
to be generated dynamically was as simple as adding one more case
to \RktSym{run{-}study} to handle callable study struct instances.
Furthermore, the combination of continuations that can close over
arbitrary Racket objects alongside the data{-}driven nature of the studies
allows us to easily create and compose studies using the full suite of
Racket{'}s facilities, including higher{-}order studies, just as we would
any other tree{-}like data structure.

To highlight that the above benefits are in no way obvious or automatic,
let us illustrate how they are absent from oTree \Autobibref{~[\hyperref[t:x28autobib_x22Daniel_Lx2e_Chenx2c_Martin_Schongerx2c_and_Chris_WickensoTreex2014An_openx2dsource_platform_for_laboratoryx2c_onlinex2c_and_field_experimentsJournal_of_Behavioral_and_Experimental_Finance_9x2c_ppx2e_88x2dx2d972016doix3a10x2e1016x2fjx2ejbefx2e2015x2e12x2e001x22x29]{\AutobibLink{1}}]},
a popular framework for economic experiments.\NoteBox{\NoteContent{Here we highlight purposefully dimensions in which oTree is
  lacking, even though oTree is clearly successful and superior to
  Congame in many respects.}} oTree
represents studies as apps that are run in a linear sequence, with
each app requiring its own folder with various files. This design
makes it hard to combine and reuse apps, not least due to difficulties
in sharing data between apps. For example, when \Scribtexttt{app2} should
be run only for participants with a high score in \Scribtexttt{app1}, then
\Scribtexttt{app1} has to store the score in a global namespace, then \Scribtexttt{app2}
looks up this score and decides whether to run or hand over to
\Scribtexttt{app3}. In Congame, \Scribtexttt{study1} can locally decide to transition
to \Scribtexttt{study2} or \Scribtexttt{study3} depending on a high or low score.
In figure~\hyperref[t:x28counter_x28x22figurex22_x22otreex2dexamplex22x29x29]{\FigureRef{5}{t:x28counter_x28x22figurex22_x22otreex2dexamplex22x29x29}} we replicate the Congame example from
figure~\hyperref[t:x28counter_x28x22figurex22_x22varx2dexamplex22x29x29]{\FigureRef{4}{t:x28counter_x28x22figurex22_x22varx2dexamplex22x29x29}} using oTree, illustrating the problem of
sharing data between apps.\NoteBox{\NoteContent{Of course, such a simple study would not
normally be split across multiple apps.}} So, while the ease of use of
oTree makes developing simple studies easy, its limitations on composing
studies and managing state make developing complex studies hard.

\begin{Herefigure}\begin{Leftfigure}\begin{FigureInside}\begin{SVerbatim}\begin{SingleColumn}\Scribtexttt{\# In settings{\hbox{\texttt{.}}}py{\hbox{\texttt{:}}}}

\Scribtexttt{SESSION{\char`\_}CONFIGS = [{\char`\{}"app{\char`\_}sequence"{\hbox{\texttt{:}}} [}

\Scribtexttt{}\mbox{\hphantom{\Scribtexttt{xx}}}\Scribtexttt{{\textquotesingle}Intro{\textquotesingle}, {\textquotesingle}Choices{\textquotesingle}, {\textquotesingle}Result{\textquotesingle}}

\Scribtexttt{]{\char`\}}]}

\Scribtexttt{PARTICIPANT{\char`\_}FIELDS = [{\textquotesingle}ok{\textquotesingle}]}

\Scribtexttt{}\mbox{\hphantom{\Scribtexttt{x}}}

\Scribtexttt{\# In Intro/IntroPage{\hbox{\texttt{.}}}html{\hbox{\texttt{:}}}}

\Scribtexttt{\# {\hbox{\texttt{.}}}{\hbox{\texttt{.}}}{\hbox{\texttt{.}}}}

\Scribtexttt{{\char`\{}{\char`\{} block content {\char`\}}{\char`\}}}

\Scribtexttt{}\mbox{\hphantom{\Scribtexttt{xx}}}\Scribtexttt{{\Stttextless}h1{\Stttextmore}Welcome to the study{\hbox{\texttt{!}}}{\Stttextless}/h1{\Stttextmore}}

\Scribtexttt{}\mbox{\hphantom{\Scribtexttt{xx}}}\Scribtexttt{{\char`\{}{\char`\{} next{\char`\_}button {\char`\}}{\char`\}}}

\Scribtexttt{{\char`\{}{\char`\{} endblock {\char`\}}{\char`\}}}

\Scribtexttt{}\mbox{\hphantom{\Scribtexttt{x}}}

\Scribtexttt{\# In Choices/{\char`\_}{\char`\_}init{\char`\_}{\char`\_}{\hbox{\texttt{.}}}py{\hbox{\texttt{:}}}}

\Scribtexttt{\# {\hbox{\texttt{.}}}{\hbox{\texttt{.}}}{\hbox{\texttt{.}}}}

\Scribtexttt{class Player(BasePlayer){\hbox{\texttt{:}}}}

\Scribtexttt{}\mbox{\hphantom{\Scribtexttt{xxxx}}}\Scribtexttt{choice = models{\hbox{\texttt{.}}}StringField(label={\textquotesingle}Choice{\hbox{\texttt{:}}}{\textquotesingle})}

\Scribtexttt{}\mbox{\hphantom{\Scribtexttt{x}}}

\Scribtexttt{class ChoicePage(Page){\hbox{\texttt{:}}}}

\Scribtexttt{}\mbox{\hphantom{\Scribtexttt{xxxx}}}\Scribtexttt{form{\char`\_}model = {\textquotesingle}player{\textquotesingle}}

\Scribtexttt{}\mbox{\hphantom{\Scribtexttt{xxxx}}}\Scribtexttt{form{\char`\_}fields = [{\textquotesingle}choice{\textquotesingle}]}

\Scribtexttt{}\mbox{\hphantom{\Scribtexttt{x}}}

\Scribtexttt{}\mbox{\hphantom{\Scribtexttt{xxxx}}}\Scribtexttt{@staticmethod}

\Scribtexttt{}\mbox{\hphantom{\Scribtexttt{xxxx}}}\Scribtexttt{def before{\char`\_}next{\char`\_}page(player, timedout){\hbox{\texttt{:}}}}

\Scribtexttt{}\mbox{\hphantom{\Scribtexttt{xxxxxxxx}}}\Scribtexttt{player{\hbox{\texttt{.}}}participant{\hbox{\texttt{.}}}ok = player{\hbox{\texttt{.}}}choice == random{\hbox{\texttt{.}}}choice([{\textquotesingle}heads{\textquotesingle}, {\textquotesingle}tails{\textquotesingle}])}

\Scribtexttt{\# {\hbox{\texttt{.}}}{\hbox{\texttt{.}}}{\hbox{\texttt{.}}}}

\Scribtexttt{}\mbox{\hphantom{\Scribtexttt{x}}}

\Scribtexttt{\# In Result/ResultPage{\hbox{\texttt{.}}}html{\hbox{\texttt{:}}}}

\Scribtexttt{\# {\hbox{\texttt{.}}}{\hbox{\texttt{.}}}{\hbox{\texttt{.}}}}

\Scribtexttt{{\char`\{}{\char`\{} block content {\char`\}}{\char`\}}}

\Scribtexttt{}\mbox{\hphantom{\Scribtexttt{xx}}}\Scribtexttt{{\char`\{}\% if player{\hbox{\texttt{.}}}participant{\hbox{\texttt{.}}}ok \%{\char`\}}}

\Scribtexttt{}\mbox{\hphantom{\Scribtexttt{xxxx}}}\Scribtexttt{{\Stttextless}p{\Stttextmore}You chose right{\hbox{\texttt{.}}}{\Stttextless}/p{\Stttextmore}}

\Scribtexttt{}\mbox{\hphantom{\Scribtexttt{xx}}}\Scribtexttt{{\char`\{}\% else \%{\char`\}}}

\Scribtexttt{}\mbox{\hphantom{\Scribtexttt{xxxx}}}\Scribtexttt{{\Stttextless}p{\Stttextmore}You chose wrong{\hbox{\texttt{.}}}{\Stttextless}/p{\Stttextmore}}

\Scribtexttt{}\mbox{\hphantom{\Scribtexttt{xx}}}\Scribtexttt{{\char`\{}\% endif \%{\char`\}}}

\Scribtexttt{{\char`\{}{\char`\{} endblock {\char`\}}{\char`\}}}\end{SingleColumn}\end{SVerbatim}\end{FigureInside}\end{Leftfigure}

\Centertext{\Legend{\FigureTarget{\label{t:x28counter_x28x22figurex22_x22otreex2dexamplex22x29x29}\textbf{Figure}~\textbf{5}. }{t:x28counter_x28x22figurex22_x22otreex2dexamplex22x29x29}A heavily edited{-}for{-}space example of the coin toss study implemented in oTree.}}\end{Herefigure}

Of course, our design could be replicated without continuations, but
continuations made this design natural and allowed us to stay flexible.

\sectionNewpage

\Ssection{Challenges of Continuations}{Challenges of Continuations}\label{t:x28part_x22challengesx22x29}

\Ssubsection{Too Few or Too Many Parameters}{Too Few or Too Many Parameters}\label{t:x28part_x22Toox5fFewx5forx5fToox5fManyx5fParametersx22x29}

To allow participants to resume a study when necessary (e.g., when
they close the browser tab and return to the website, after their
continuations expire, or after a server re{-}deployment), Congame
persists the participant{'}s position: the fully{-}qualified path to the
node they reached within the study tree as represented as a list of
ids. In memory, this position is tracked using dynamic variables
(\emph{parameters}\Autobibref{~[\hyperref[t:x28autobib_x22Matthew_Flattx2c_Gang_Yux2c_Robert_Bruce_Findlerx2c_and_Matthias_FelleisenAdding_delimited_and_composable_control_to_a_production_programming_environmentACM_SIGPLAN_Notices_42x2c_ppx2e_165x2dx2d1762007doix3a10x2e1145x2f1291220x2e1291178x22x29]{\AutobibLink{4}}]} in Racket
parlance). In specific cases, continuations interact with parameters in
surprising ways.

When a continuation URL is visited and the continuation is restored, it
is run in a fresh Racket thread. Racket threads inherit the parameters
of their parent threads. That is, if a parameter is set to one value
in the parent, it will be set to the same value in the child thread.
When \RktSym{parameterize} is used to change the value of a parameter
for a particular block of code, instead of storing each new parameter
value in a thread cell individually, a parameterization object is
extended\NoteBox{\NoteContent{\href{https://github.com/racket/racket/issues/4216}{\Snolinkurl{https://github.com/racket/racket/issues/4216}}}} to include the new values of the changed
parameters. As a consequence, when \RktSym{parameterize} is used within
the dynamic extent of a continuation and that continuation is later
restored in a thread, more parameters than might be expected may end up
being restored, because the aforementioned extended parameterization
object is installed alongside it.

Figure~\hyperref[t:x28counter_x28x22figurex22_x22challengex2d1x22x29x29]{\FigureRef{6}{t:x28counter_x28x22figurex22_x22challengex2d1x22x29x29}} shows an example of this issue. When run, the program
in figure 4 displays {``}a b{''}; but, since the continuation is captured
up to a prompt that resides within the outer \RktSym{parameterize} form setting
the parameter \RktSym{a}, we had initially expected to see {``}\#f b{''}. Removing
the inner use of \RktSym{parameterize} causes the program to display {``}\#f \#f{''}.

\begin{Herefigure}\begin{Centerfigure}\begin{FigureInside}\begin{RktBlk}\begin{SingleColumn}\RktPn{(}\RktSym{define}\mbox{\hphantom{\Scribtexttt{x}}}\RktSym{a}\mbox{\hphantom{\Scribtexttt{x}}}\RktPn{(}\RktSym{make{-}parameter}\mbox{\hphantom{\Scribtexttt{x}}}\RktVal{\#f}\RktPn{)}\RktPn{)}

\RktPn{(}\RktSym{define}\mbox{\hphantom{\Scribtexttt{x}}}\RktSym{b}\mbox{\hphantom{\Scribtexttt{x}}}\RktPn{(}\RktSym{make{-}parameter}\mbox{\hphantom{\Scribtexttt{x}}}\RktVal{\#f}\RktPn{)}\RktPn{)}

\RktPn{(}\RktSym{define}\mbox{\hphantom{\Scribtexttt{x}}}\RktSym{tag}\mbox{\hphantom{\Scribtexttt{x}}}\RktPn{(}\RktSym{make{-}continuation{-}prompt{-}tag}\RktPn{)}\RktPn{)}

\RktPn{(}\RktSym{define}\mbox{\hphantom{\Scribtexttt{x}}}\RktSym{k}

\mbox{\hphantom{\Scribtexttt{xx}}}\RktPn{(}\RktSym{parameterize}\mbox{\hphantom{\Scribtexttt{x}}}\RktPn{(}\RktPn{[}\RktSym{a}\mbox{\hphantom{\Scribtexttt{x}}}\RktVal{{\textquotesingle}}\RktVal{a}\RktPn{]}\RktPn{)}

\mbox{\hphantom{\Scribtexttt{xxxx}}}\RktPn{(}\RktSym{call{-}with{-}continuation{-}prompt}

\mbox{\hphantom{\Scribtexttt{xxxxx}}}\RktPn{(}\RktSym{lambda}\mbox{\hphantom{\Scribtexttt{x}}}\RktPn{(}\RktPn{)}

\mbox{\hphantom{\Scribtexttt{xxxxxxx}}}\RktPn{(}\RktSym{parameterize}\mbox{\hphantom{\Scribtexttt{x}}}\RktPn{(}\RktPn{[}\RktSym{b}\mbox{\hphantom{\Scribtexttt{x}}}\RktVal{{\textquotesingle}}\RktVal{b}\RktPn{]}\RktPn{)}

\mbox{\hphantom{\Scribtexttt{xxxxxxxxx}}}\RktPn{(}\RktPn{(}\RktSym{call{-}with{-}current{-}continuation}

\mbox{\hphantom{\Scribtexttt{xxxxxxxxxxx}}}\RktPn{(}\RktSym{$\lambda$}\mbox{\hphantom{\Scribtexttt{x}}}\RktPn{(}\RktSym{k}\RktPn{)}\mbox{\hphantom{\Scribtexttt{x}}}\RktPn{(}\RktSym{thunk}\mbox{\hphantom{\Scribtexttt{x}}}\RktSym{k}\RktPn{)}\RktPn{)}

\mbox{\hphantom{\Scribtexttt{xxxxxxxxxxx}}}\RktSym{tag}\RktPn{)}\RktPn{)}\RktPn{)}\RktPn{)}

\mbox{\hphantom{\Scribtexttt{xxxxx}}}\RktSym{tag}\RktPn{)}\RktPn{)}\RktPn{)}

\RktPn{(}\RktSym{call{-}with{-}continuation{-}prompt}

\mbox{\hphantom{\Scribtexttt{x}}}\RktPn{(}\RktSym{lambda}\mbox{\hphantom{\Scribtexttt{x}}}\RktPn{(}\RktPn{)}

\mbox{\hphantom{\Scribtexttt{xxx}}}\RktPn{(}\RktSym{k}\mbox{\hphantom{\Scribtexttt{x}}}\RktPn{(}\RktSym{lambda}\mbox{\hphantom{\Scribtexttt{x}}}\RktPn{(}\RktPn{)}

\mbox{\hphantom{\Scribtexttt{xxxxxxxx}}}\RktPn{(}\RktSym{printf}\mbox{\hphantom{\Scribtexttt{x}}}\RktVal{"{\textasciitilde}s {\textasciitilde}s{\textasciitilde}n"}\mbox{\hphantom{\Scribtexttt{x}}}\RktPn{(}\RktSym{a}\RktPn{)}\mbox{\hphantom{\Scribtexttt{x}}}\RktPn{(}\RktSym{b}\RktPn{)}\RktPn{)}\RktPn{)}\RktPn{)}\RktPn{)}

\mbox{\hphantom{\Scribtexttt{x}}}\RktSym{tag}\RktPn{)}\end{SingleColumn}\end{RktBlk}\end{FigureInside}\end{Centerfigure}

\Centertext{\Legend{\FigureTarget{\label{t:x28counter_x28x22figurex22_x22challengex2d1x22x29x29}\textbf{Figure}~\textbf{6}. }{t:x28counter_x28x22figurex22_x22challengex2d1x22x29x29}An example of the parameter revival issue.}}\end{Herefigure}

On the opposite side of the coin, because the Racket web server restores
continuations in fresh threads, it is also possible to {``}lose{''} changes
to a parameter when using direct assignment. Directly assigning a
parameter in a thread records the change to the parameter in a thread
cell, without affecting the current parameterization. Originally,
we had used direct assignment to update the participant{'}s position,
which caused the parameter to reset at the boundary between steps.
Then, we switched to explicit uses of \RktSym{parameterize}, which
extends the parameterization such that the changes are available in the
restored continuation, as in the previous example. However, this was not
foolproof since whether or not a parameterization is extended depends
on where the \RktSym{parameterize} is situated in the dynamic extent of
the delimited continuation: if it is before the prompt, the extension is
not visible, otherwise it is. Finally, we settled on manually passing
around the parameterization between steps to have full control over what
values the parameters we depend on have at any time. We have not yet
experimented with using continuation marks \Autobibref{~[\hyperref[t:x28autobib_x22John_Clementsx2c_Matthew_Flattx2c_and_Matthias_FelleisenModeling_an_Algebraic_StepperLecture_Notes_in_Computer_Science_2028x2c_ppx2e_320x2dx2d3342001doix3a10x2e1007x2f3x2d540x2d45309x2d1x5f21x22x29]{\AutobibLink{2}}]} directly.

\begin{Herefigure}\begin{Centerfigure}\begin{FigureInside}\begin{RktBlk}\begin{SingleColumn}\RktPn{(}\RktSym{define}\mbox{\hphantom{\Scribtexttt{x}}}\RktSym{p}\mbox{\hphantom{\Scribtexttt{x}}}\RktPn{(}\RktSym{make{-}parameter}\mbox{\hphantom{\Scribtexttt{x}}}\RktVal{\#f}\RktPn{)}\RktPn{)}

\RktPn{(}\RktSym{define}\mbox{\hphantom{\Scribtexttt{x}}}\RktSym{tag}\mbox{\hphantom{\Scribtexttt{x}}}\RktPn{(}\RktSym{make{-}continuation{-}prompt{-}tag}\RktPn{)}\RktPn{)}

\RktPn{(}\RktSym{define}\mbox{\hphantom{\Scribtexttt{x}}}\RktSym{k{-}ch}\mbox{\hphantom{\Scribtexttt{x}}}\RktPn{(}\RktSym{make{-}channel}\RktPn{)}\RktPn{)}

\RktPn{(}\RktSym{void}

\mbox{\hphantom{\Scribtexttt{x}}}\RktPn{(}\RktSym{thread}

\mbox{\hphantom{\Scribtexttt{xx}}}\RktPn{(}\RktSym{lambda}\mbox{\hphantom{\Scribtexttt{x}}}\RktPn{(}\RktPn{)}

\mbox{\hphantom{\Scribtexttt{xxxx}}}\RktPn{(}\RktSym{call{-}with{-}continuation{-}prompt}

\mbox{\hphantom{\Scribtexttt{xxxxx}}}\RktPn{(}\RktSym{lambda}\mbox{\hphantom{\Scribtexttt{x}}}\RktPn{(}\RktPn{)}

\mbox{\hphantom{\Scribtexttt{xxxxxxx}}}\RktPn{(}\RktSym{parameterize}\mbox{\hphantom{\Scribtexttt{x}}}\RktPn{(}\RktPn{[}\RktSym{p}\mbox{\hphantom{\Scribtexttt{x}}}\RktVal{{\textquotesingle}}\RktVal{p1}\RktPn{]}\RktPn{)}

\mbox{\hphantom{\Scribtexttt{xxxxxxxxx}}}\RktPn{(}\RktSym{p}\mbox{\hphantom{\Scribtexttt{x}}}\RktVal{{\textquotesingle}}\RktVal{p2}\RktPn{)}

\mbox{\hphantom{\Scribtexttt{xxxxxxxxx}}}\RktPn{(}\RktPn{(}\RktSym{call{-}with{-}current{-}continuation}

\mbox{\hphantom{\Scribtexttt{xxxxxxxxxxx}}}\RktPn{(}\RktSym{lambda}\mbox{\hphantom{\Scribtexttt{x}}}\RktPn{(}\RktSym{k}\RktPn{)}

\mbox{\hphantom{\Scribtexttt{xxxxxxxxxxxxx}}}\RktPn{(}\RktSym{thunk}\mbox{\hphantom{\Scribtexttt{x}}}\RktPn{(}\RktSym{channel{-}put}\mbox{\hphantom{\Scribtexttt{x}}}\RktSym{k{-}ch}\mbox{\hphantom{\Scribtexttt{x}}}\RktSym{k}\RktPn{)}\RktPn{)}\RktPn{)}

\mbox{\hphantom{\Scribtexttt{xxxxxxxxxxx}}}\RktSym{tag}\RktPn{)}\RktPn{)}\RktPn{)}\RktPn{)}

\mbox{\hphantom{\Scribtexttt{xxxxx}}}\RktSym{tag}\RktPn{)}\RktPn{)}\RktPn{)}\RktPn{)}

\RktPn{(}\RktSym{thread{-}wait}

\mbox{\hphantom{\Scribtexttt{x}}}\RktPn{(}\RktSym{thread}

\mbox{\hphantom{\Scribtexttt{xx}}}\RktPn{(}\RktSym{lambda}\mbox{\hphantom{\Scribtexttt{x}}}\RktPn{(}\RktPn{)}

\mbox{\hphantom{\Scribtexttt{xxxx}}}\RktPn{(}\RktSym{define}\mbox{\hphantom{\Scribtexttt{x}}}\RktSym{k}\mbox{\hphantom{\Scribtexttt{x}}}\RktPn{(}\RktSym{channel{-}get}\mbox{\hphantom{\Scribtexttt{x}}}\RktSym{k{-}ch}\RktPn{)}\RktPn{)}

\mbox{\hphantom{\Scribtexttt{xxxx}}}\RktPn{(}\RktSym{call{-}with{-}continuation{-}prompt}

\mbox{\hphantom{\Scribtexttt{xxxxx}}}\RktPn{(}\RktSym{lambda}\mbox{\hphantom{\Scribtexttt{x}}}\RktPn{(}\RktPn{)}

\mbox{\hphantom{\Scribtexttt{xxxxxxx}}}\RktPn{(}\RktSym{k}\mbox{\hphantom{\Scribtexttt{x}}}\RktPn{(}\RktSym{$\lambda$}\mbox{\hphantom{\Scribtexttt{x}}}\RktPn{(}\RktPn{)}\mbox{\hphantom{\Scribtexttt{x}}}\RktPn{(}\RktSym{printf}\mbox{\hphantom{\Scribtexttt{x}}}\RktVal{"{\textasciitilde}s{\textasciitilde}n"}\mbox{\hphantom{\Scribtexttt{x}}}\RktPn{(}\RktSym{p}\RktPn{)}\RktPn{)}\RktPn{)}\RktPn{)}\RktPn{)}

\mbox{\hphantom{\Scribtexttt{xxxxx}}}\RktSym{tag}\RktPn{)}\RktPn{)}\RktPn{)}\RktPn{)}\end{SingleColumn}\end{RktBlk}\end{FigureInside}\end{Centerfigure}

\Centertext{\Legend{\FigureTarget{\label{t:x28counter_x28x22figurex22_x22challengex2d2x22x29x29}\textbf{Figure}~\textbf{7}. }{t:x28counter_x28x22figurex22_x22challengex2d2x22x29x29}An example of the parameter loss issue.}}\end{Herefigure}

Figure~\hyperref[t:x28counter_x28x22figurex22_x22challengex2d2x22x29x29]{\FigureRef{7}{t:x28counter_x28x22figurex22_x22challengex2d2x22x29x29}} demonstrates the parameter loss issue. When the
continuation from the first thread is restored in the second, the direct
assignment to the parameter is lost and the program displays {``}p1{''}.

\Ssubsection{Debugging}{Debugging}\label{t:x28part_x22debuggingx22x29}

Debugging memory leaks in the presence of continuations is tricky. We
had a set of small bugs in different areas of the system that were
composing together to form a larger bug which led to massive memory
leaks under load.

First, our error reporting library was setting up exception
handlers in its inner data collection loop, making the loop no
longer tail{-}recursive. Second, our own middleware to configure the
aforementioned error reporting library was accidentally creating a new
instance of the error reporter per request, instead of reusing a single
one, meaning that for every new request we would spin up a new data
collection thread with a non{-}tail{-}recursive inner loop. Finally, we were
using composable continuations to implement a special type of return
from a sub{-}study to its parent, so when a participant continued a study
at this boundary between parent and sub{-}study, we would see an increase
in memory usage from stacking the composable continuations on top of
each other.

\begin{Herefigure}\begin{Centerfigure}\begin{FigureInside}\includegraphics[scale=0.33]{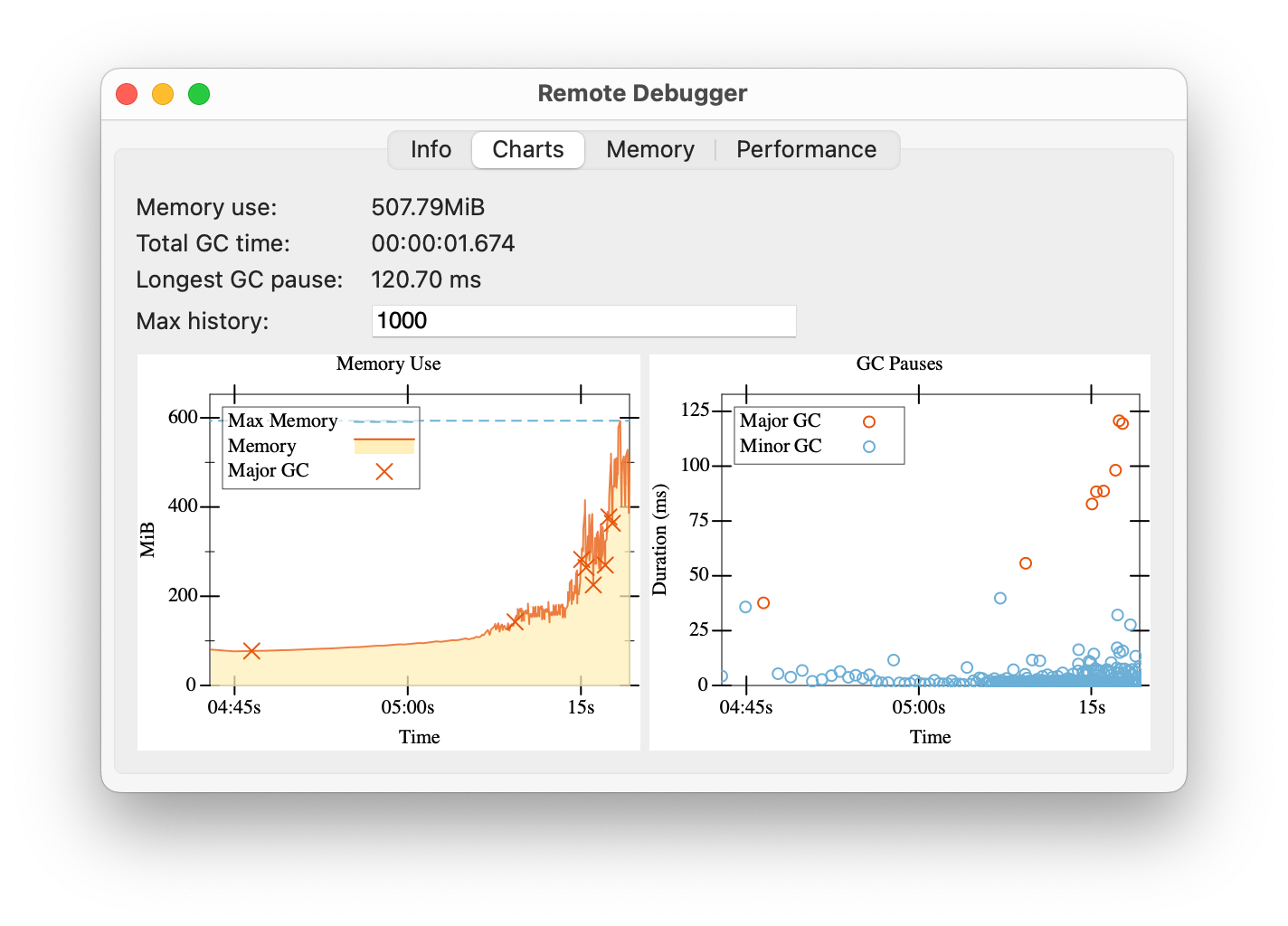}
\includegraphics[scale=0.33]{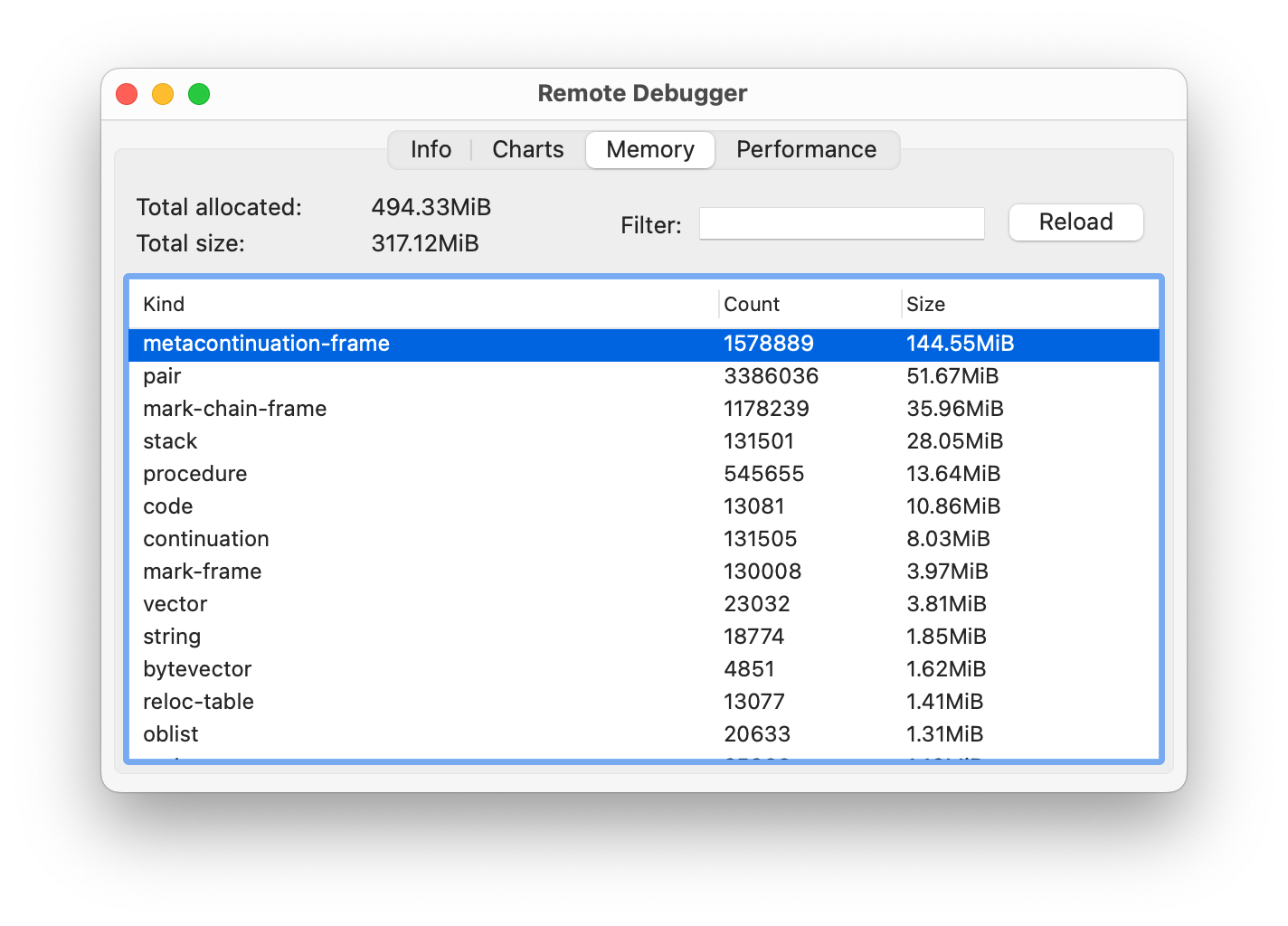}\end{FigureInside}\end{Centerfigure}

\Centertext{\Legend{\FigureTarget{\label{t:x28counter_x28x22figurex22_x22remotex2ddebuggerx22x29x29}\textbf{Figure}~\textbf{8}. }{t:x28counter_x28x22figurex22_x22remotex2ddebuggerx22x29x29}A remote debugger for Racket showing a spike in memory usage and the values that are taking up that memory.}}\end{Herefigure}

Figure~\hyperref[t:x28counter_x28x22figurex22_x22remotex2ddebuggerx22x29x29]{\FigureRef{8}{t:x28counter_x28x22figurex22_x22remotex2ddebuggerx22x29x29}} shows what this type of issue looks like
when visualized using dbg \Autobibref{~[\hyperref[t:x28autobib_x22Bogdan_Popadbg2021httpsx3ax2fx2fgithubx2ecomx2fBogdanpx2fracketx2ddbgx22x29]{\AutobibLink{6}}]}, a remote debugging tool for
Racket. We can see memory use grow exponentially and that this stems
from allocating a lot of {``}metacontinuation{-}frame{''} values. This drew
our attention to our use of composable continuations, which we promptly
changed to delimited{-}but{-}not{-}composable continuations, since we didn{'}t
actually need composable continuations for our purposes. Our use of
composable continuations amplified the other two bugs, and this change
seemed to fix the issue by drastically reducing the effect of the memory
leak. In a way, this fix gave us a false sense of security, since the
other two problems were still lurking, so we were surprised to later run
into the same problem again. Eventually, we were able to find the root
problems and fix them.

\sectionNewpage

\Ssection{Reflections \& Recommendations}{Reflections \& Recommendations}\label{t:x28part_x22reflectionsx22x29}

Continuations allow us to write web code in a direct style, simplifying
the job of embedding domain specific languages in a web context.
Without leveraging continuations, the study harness introduced in
\ChapRef{\SectionNumberLink{t:x28part_x22minimalx22x29}{2}}{Mini Congame} would have been a lot more cumbersome to implement
than a straightforward depth{-}first traversal of a tree. The issues
we encountered were primarily due to unexpected interactions
between dynamic variables and delimited continuations. We recommend
that functional architects avoid combining dynamic variables and
continuations in their systems where possible, or do so with care,
while taking into account the issues presented in \ChapRef{\SectionNumberLink{t:x28part_x22challengesx22x29}{4}}{Challenges of Continuations}.
If we were to rewrite Congame today, we would explicitly pass around
a context object containing the data we need between steps instead of
using dynamic variables. We also recommend that functional architects
carefully consider whether they need composable \emph{and} delimited
continuations, or whether delimited alone will suffice.

To aid debugging, languages should provide tooling to allow
continuations to be inspected at runtime. That way, when encountering
issues such as the one presented in \SecRef{\SectionNumberLink{t:x28part_x22debuggingx22x29}{4.2}}{Debugging}, programmers
would have an easier time finding the source of memory leaks. For
example, in the case of Racket{'}s {``}metacontinuation{-}frame{''} values, it
would be helpful if those values were inspectable to determine their
source location and what other objects they hold references to.

We believe continuations are the right abstraction for implementing
interactive systems as targeted by Congame, such as surveys and market
games, as well as any other system that requires some computation to
be suspended until the user takes action (e.g. shopping carts, or
simulations where part of the computation is delegated to another black
box, etc.). For our use case, where backtracking via the browser{'}s
{``}Back{''} button is undesirable, multi{-}shot continuations are not
required but, in other applications, they may be useful. In that sense,
in a language without continuations, coroutines would likely provide
us with the same benefits, but would not be as good a fit for use
cases where the possibility of branching the user{'}s progress through an
interaction is a desirable feature (e.g. having the ability to open
a separate tab to take a different path through a study tree). Other
approaches, such as regular web programming with manual routing or a
weaker form of {``}continuations{''} where URLs get mapped to closures,
would not permit us to implement the core study harness in such a direct
and simple way.

\begin{acks}

We would like to thank the anonymous reviewers for their comments
and suggestions. This work was partially funded by Central European
University Private University.

\end{acks}

\sectionNewpage

\Ssectionstarx{References}{References}\label{t:x28part_x22docx2dbibliographyx22x29}

\begin{bigtabular}{@{\bigtableleftpad}l@{}l@{}}
\hbox{\Autocolbibnumber{[1]}} &
\hbox{\Autobibtarget{\label{t:x28autobib_x22Daniel_Lx2e_Chenx2c_Martin_Schongerx2c_and_Chris_WickensoTreex2014An_openx2dsource_platform_for_laboratoryx2c_onlinex2c_and_field_experimentsJournal_of_Behavioral_and_Experimental_Finance_9x2c_ppx2e_88x2dx2d972016doix3a10x2e1016x2fjx2ejbefx2e2015x2e12x2e001x22x29}\Autocolbibentry{Daniel L. Chen, Martin Schonger, and Chris Wickens. oTree{---}An open{-}source platform for laboratory, online, and field experiments. \textit{Journal of Behavioral and Experimental Finance} 9, pp. 88{--}97, 2016. \pseudodoi{doi:\href{https://doi.org/10.1016/j.jbef.2015.12.001}{10{\hbox{\texttt{.}}}1016/j{\hbox{\texttt{.}}}jbef{\hbox{\texttt{.}}}2015{\hbox{\texttt{.}}}12{\hbox{\texttt{.}}}001}}}}} \\
\hbox{\Autocolbibnumber{[2]}} &
\hbox{\Autobibtarget{\label{t:x28autobib_x22John_Clementsx2c_Matthew_Flattx2c_and_Matthias_FelleisenModeling_an_Algebraic_StepperLecture_Notes_in_Computer_Science_2028x2c_ppx2e_320x2dx2d3342001doix3a10x2e1007x2f3x2d540x2d45309x2d1x5f21x22x29}\Autocolbibentry{John Clements, Matthew Flatt, and Matthias Felleisen. Modeling an Algebraic Stepper. \textit{Lecture Notes in Computer Science} 2028, pp. 320{--}334, 2001. \pseudodoi{doi:\href{https://doi.org/10.1007/3-540-45309-1_21}{10{\hbox{\texttt{.}}}1007/3{-}540{-}45309{-}1{\char`\_}21}}}}} \\
\hbox{\Autocolbibnumber{[3]}} &
\hbox{\Autobibtarget{\label{t:x28autobib_x22Matthew_Flatt_and_PLTReferencex3a_RacketPLT_Design_Incx2ex2c_PLTx2dTRx2d2010x2d12010httpsx3ax2fx2fracketx2dlangx2eorgx2ftr1x2fx22x29}\Autocolbibentry{Matthew Flatt and PLT. Reference: Racket. PLT Design Inc., PLT{-}TR{-}2010{-}1, 2010. \href{https://racket-lang.org/tr1/}{\Snolinkurl{https://racket-lang.org/tr1/}}}}} \\
\hbox{\Autocolbibnumber{[4]}} &
\hbox{\Autobibtarget{\label{t:x28autobib_x22Matthew_Flattx2c_Gang_Yux2c_Robert_Bruce_Findlerx2c_and_Matthias_FelleisenAdding_delimited_and_composable_control_to_a_production_programming_environmentACM_SIGPLAN_Notices_42x2c_ppx2e_165x2dx2d1762007doix3a10x2e1145x2f1291220x2e1291178x22x29}\Autocolbibentry{Matthew Flatt, Gang Yu, Robert Bruce Findler, and Matthias Felleisen. Adding delimited and composable control to a production programming environment. \textit{ACM SIGPLAN Notices} 42, pp. 165{--}176, 2007. \pseudodoi{doi:\href{https://doi.org/10.1145/1291220.1291178}{10{\hbox{\texttt{.}}}1145/1291220{\hbox{\texttt{.}}}1291178}}}}} \\
\hbox{\Autocolbibnumber{[5]}} &
\hbox{\Autobibtarget{\label{t:x28autobib_x22Shriram_Krishnamurthix2c_Peter_Walton_Hopkinsx2c_Jay_McCarthyx2c_Paul_Tx2e_Graunkex2c_Greg_Pettyjohnx2c_and_Matthias_FelleisenImplementation_and_use_of_the_PLT_scheme_Web_serverHigherx2dOrder_and_Symbolic_Computation_20x2c_ppx2e_431x2dx2d4602007doix3a10x2e1007x2fs10990x2d007x2d9008x2dyx22x29}\Autocolbibentry{Shriram Krishnamurthi, Peter Walton Hopkins, Jay McCarthy, Paul T. Graunke, Greg Pettyjohn, and Matthias Felleisen. Implementation and use of the PLT scheme Web server. \textit{Higher{-}Order and Symbolic Computation} 20, pp. 431{--}460, 2007. \pseudodoi{doi:\href{https://doi.org/10.1007/s10990-007-9008-y}{10{\hbox{\texttt{.}}}1007/s10990{-}007{-}9008{-}y}}}}} \\
\hbox{\Autocolbibnumber{[6]}} &
\hbox{\Autobibtarget{\label{t:x28autobib_x22Bogdan_Popadbg2021httpsx3ax2fx2fgithubx2ecomx2fBogdanpx2fracketx2ddbgx22x29}\Autocolbibentry{Bogdan Popa. dbg. 2021. \href{https://github.com/Bogdanp/racket-dbg}{\Snolinkurl{https://github.com/Bogdanp/racket-dbg}}}}} \\
\hbox{\Autocolbibnumber{[7]}} &
\hbox{\Autobibtarget{\label{t:x28autobib_x22Christian_QueinnecInverting_back_the_inversion_of_control_orx2c_continuations_versus_pagex2dcentric_programmingACM_SIGPLAN_Notices_38x2c_ppx2e_57x2dx2d642003doix3a10x2e1145x2f772970x2e772977x22x29}\Autocolbibentry{Christian Queinnec. Inverting back the inversion of control or, continuations versus page{-}centric programming. \textit{ACM SIGPLAN Notices} 38, pp. 57{--}64, 2003. \pseudodoi{doi:\href{https://doi.org/10.1145/772970.772977}{10{\hbox{\texttt{.}}}1145/772970{\hbox{\texttt{.}}}772977}}}}}\end{bigtabular}

\postDoc
\end{document}